\documentstyle[german,epsfig,sprocl]{article}

\def\Journal#1#2#3#4{{#1} {\bf #2}, #3 (#4)}

\def\NIMA{{\em Nucl. Instrum. Methods} A}

\def\be{\begin{equation}}
\def\ee{\end{equation}}
\def\bea{\begin{eqnarray}}
\def\eea{\end{eqnarray}}

\begin{document}
\bibliographystyle{unsrt}

\title{MSGC Development for HERA-B}

\author{ B. SCHMIDT }

\address{ Physikalisches Institut, University of Heidelberg, 
Philosophenweg 12,
\newline D-69120 Heidelberg, Germany}


\maketitle\abstracts{The Inner Tracker System of the HERA-B 
experiment at DESY is build by \mbox{groups \cite{names}} at the 
Universities of Heidelberg, Siegen and Z"urich. The system consists 
of 184 Microstrip-Gas-Chambers (MSGC) with a total number of 147 456 
electronic channels. The detectors have to cope with particle fluxes 
up to 25 kHz mm$-2$ and to tolerate radiation doses of 1 Mrad per 
year. During the development of these chambers it was found that 
conventional MSGC, operated in intense hadronic fluxes, are rapidly 
destroyed due to the phenomenon of induced discharges. The 
introduction of a Gas Electron Multiplier (GEM) as preamplification 
structure offers the possibility to build robust and reliable 
detectors allowing for ample gain reserve in the hostile environment 
of HERA-B.}

\section{Introduction}
The goal of HERA-B is to establish CP violation in B-decays 
\cite{tdr}. As source of the B mesons, an internal wire target in the 
800 MeV proton beam line of HERA is used. Thus HERA-B is a fixed 
target experiment of type forward spectrometer. The specific 
requirements of the detector elements are due to the enormous amount 
of 4 $\times 10^{14}$ events which have to be recorded per year and 
from which about 1000 of the rare decays into J/$\Psi K_s$ have to be 
extracted. 

This means that HERA-B has to run at a mean interaction rate of 
\mbox{40 MHz}. Four overlayed events per bunch crossing every 96 nsec 
with a mean number of about 200 tracks have to be recorded and 
analyzed. The events are strongly boosted in the forward direction 
and about half of the tracks have to be followed in the Inner Tracker 
system, covering the area from 5-30 cm around the beam pipe. As 
solution for the inner tracker detector elements, 
Microstrip-Gas-Chambers \cite{oed} (MSGC) were foreseen and are, in 
the new version as  \mbox{GEM-MSGC \cite{sauli},} the now approved 
and accepted technology. 

\section{The parameters of the HERA-B Inner Tracker}
The task of the Inner Tracker detectors is twofold since they have to 
deliver tracking information and to provide fast signals for the 
first level trigger simultaneously. The tracking requirements are 
rather moderate, 100 $\mu$m resolution in the bending plane and 1 mm 
resolution parallel to the B-field are sufficient. Part of the 
detectors have to operate in a magnetic field of 0,85 T. 

The tracking information is obtained in the conventional way using 
analog signals from individual strips. The second coordinate is 
measured from stereo layers under small angles ($5^o$) with respect 
to the magnetic field. The small stereo angles are necessary to keep 
the combinatorial background reasonably small.

For the trigger signals 4 neighbouring strips are electrically ored 
to reduce the number of trigger channels while keeping the occupancy 
at the 10-15 $\%$ level. To enhance the trigger efficiency, all 
trigger layers are double layers of identical orientation. The 
chambers have to cope with particle fluxes up to $10^4$ particles per 
mm$^2$ and second, accumulating to a total radiation dose of up to 1 
Mrad per year. The fundamental question of MSGC development and 
design was to find a solution which guarantees a stable operation 
with high gain and high efficiency over several years of HERA-B 
operation. 

The demand for fast trigger signals puts specific boundary conditions 
to the operating parameters of the chambers. The pulse shaping has to 
be done in a way that pile up and resulting fake triggers in 
successive bunch crossings are kept at tolerable low level. A 
possible shaping as it could be realized with the foreseen Helix 
pre-amplifier is shown as inset in figure 1. 

Taking the charge collection time, the ion movement and the amplifier 
shaping into account it becomes apparent that the effictive number of 
electrons contributing to the signal at optimal sampling time (95 
nsec after bunch crossing) has a broad distribution with a mean 
number of only 15.6 electrons (Figure 1). Taking 7000 electrons as a 
realistic threshold for a strip capacity of 40 pF (25 cm long strips) 
it turns out that a minimal gas gain above 2000 is required to exceed 
the 95 $\%$ efficiency level.

Simultaneously the number of fake triggers due to pulse fluctuations 
into the subsequent bunch crossing does not exceed 25 $\%$. (Figure 
2). It should be mentioned that the rather delicate balance between 
efficiency and fake trigger rate requires a sufficiently high 
uniformity of gas gain, electronic amplification and trigger 
threshold.

To summarize the scope of the project, a few numbers should be given: 
\\
The Inner Tracker of HERA-B will consist of 46 detector planes 
grouped into 10 superlayers consisting of 184 MSGC and covering a 
total area of about \mbox{16 m$^2$}. 147456 channels are equiped with 
analog read-out and 9984 fast trigger signals provided for the first 
level trigger.
 
The entire project comprising the development of the chambers, the 
infra\-structure and the read-out electronics is realized by groups 
at the universities of Heidelberg, Siegen and Z"urich.
\section{Early findings and solutions}
At the time of the technical design report (1994) the development of 
MSGC for high rate operation seemed to be settled. 

Unfortunately it turned out rather quickly, that the misgivings of 
various people were justified and the available MSGC technology did 
not allow a stable operation at high rates on a time scale required 
for HERA-B. Even observing the strongest precautions for gas quality 
and chamber material, the detectors died suddenly after collecting 
only moderate amounts of total charge. Typical examples \cite{hott} 
are given in figure 3. Especially operation at the required high 
gains turned out to be very short lived. 

The reason for that delicacy comes from the fact, that in a MSGC, 
produced on bare glass plates with ionic conductivity, the electric 
field close to the surface strongly depends on the distribution of 
the ions in the glass. Under the combined action of radiation and 
electric fields this distribution is disturbed in a way, that the gas 
gain finally breaks together and stable operation is not longer 
possible. To get rid of accumulating surface charges only electronic 
conductivity of the substrate promise a steady and uniform operation. 
Electronically conductive glass is not available in the required 
large plates and ruled out due to its very short radiation length.

From a variety of efforts to find an adequate stable surface coating 
of the glass \cite{dlc2} the diamond like coating (DLC) with 
amorphous carbon using a plasma CVD process  \cite{dlc1,dlc2} turned 
out to be the most promising solution. The DLC coatings could be well 
adjusted to the required surface conductivity of $10^{14}-10^{16}$ 
$\Omega$ per square and offer reasonable surface quality for the MSGC 
production process. 

The successful reproduction of the pioneering results of reference 6 
showing stable operation even under strong conditions and high gas 
gains, mark the first milestone of MSGC development for HERA-B 
(Figure 4). 

\section{The phenomenon of induced discharges}
After a variety of successful tests in the lab we went on a first 
hadronic beam test with fullsize HERA-B chambers at PSI in April 
1996. There we made the painful discovery that testing with x-rays in 
the lab is not sufficient to simulate the conditions of high 
intensity hadronic beams. As a new phenomenon we observed frequent 
anode-cathode discharges visible as short current spikes under 
conditions where the chambers operated with x-rays of similar 
intensity completely quietly and unconspecious. After a few hours 
operation, an alarming number of anodes was broken  \cite{vis}. 

Visual inspection of the chambers showed an enormous number of 
correlated marks in anodes and cathodes in the area where the chamber 
had been exposed to the beam (Figure 5). Part of them had led to 
anode breaks. The fact that comparable phenomena had never been 
observed with x-rays or with $\beta$-sources but only in the hadronic 
beam made it fair to assume that the discharges are induced by heavy 
ionizing particles. In a hadronic beam these heavy ionizing particles 
can be easily produced by nuclear reactions in the MSGC substrate. 

To simulate these conditions and to verify the assumption we 
introduced a gaseous $\alpha$-source in the chamber by flushing the 
counting gas through a cylinder containing thorium oxyde  \cite{vis}. 
The Rn-220 emanating from the thorium powder is transported to the 
chamber with the gas stream and decays with the half life of 55 s 
emitting 6,3 MeV $\alpha$-particles. In fact the phenomenon of 
induced discharges could be reproduced with all aspects: current 
spikes, anode-cathode marks and finally broken anodes. 

A concluding test of the chambers at the HERA-B beam line made clear 
that under nominal conditions of a gas gain around 3000 the chambers 
are destroyed within a few hours. In the following we concentrated 
all efforts to study this phenomenon of induced discharges in more 
detail in view of how to avoid it and to extend the lifetime of the 
chambers. 

From the bulk of results  \cite{cbr} the most important will be 
summarized here. \\

Figure 6 shows a strong dependence of the discharge rate on the 
cathode potential at constant gas gain. The obvious recommendation 
from this is to operate the chamber at the highest possible drift 
field. For a constant cathode potential no dependence of the 
discharge rate on the drift field could be observed, even if the 
drift field was reversed (Figure 7). This could be taken as a strong 
evidence that the triggering high primary ionization has to be 
produced very close to the MSGC surface. 

With special emphasis we checked the possibility to reduce the 
proneness to induce discharges by changing the composition and the 
nature of the counting gas. For various reasons (aging, drift 
velocity, diffusion, density of primary ionization) Ar/DME 50:50 was 
chosen at the optimal composition till then. Within the scope of 
these investigations we studied different Ar/DME mixtures as well as 
mixtures of Ne/DME, Ar/CO2, Ar/DEE with water, alcohol, methylal and 
ammonia as additives. 

Unfortunately, none of these mixtures turned out to be significantly 
less prone to induce discharges than the standard Ar/DME mixture. For 
identical gas gain all mixtures exhibited comparable discharge rates 
under the influence of the gasous $\alpha$-source. 

As a typical example figure 8 shows the discharge rate in two 
different Ar/DME mixtures: \\

For identical gain and drift field the discharge rate is completely 
independent of the gas composition. 
\\[0.3cm]
Another interesting observation is the fact that MSGC on uncoated and 
resistively coated substrates show quite different discharge rates 
under otherwise identical conditions. If we take the pulse height to 
define the operation point of the chamber, the discharge rate on 
uncoated plates is reduced by a gain dependend factor between 10 and 
20 compared to diamond coated MSGC's (Figure 9). 

The reason for that becomes obvious from figure 10, where for both 
cases the strength of the electric field between anode and cathode is 
shown for various heights above the MSGC plane.\footnote{The 
calculations have been done using the programme ACE by curtesy of ABB 
Cooperation} For the coated plate, a constant surface restistivity 
was assumed whereas the field for the uncoated plate was calculated 
using infinite surface resistivity. 

Since the actual ion distribution is not known, the latter has to be 
taken as an approximation to the real conditions. Whereas for 
uncoated plates the anode-cathode field is strongly peaked close to 
the electrodes with a broad regime of strongly reduced field in 
between, coated plates have an almost uniform field between anode and 
cathode. The low field regime on uncoated plates efficiently stops 
the evolution of streamers and suppresses the tendency for induced 
discharges.

Summarizing we have to conclude, that induced discharges are an 
intrinsic problem of the MSGC geometry and principal; gain and 
discharge rate are strongly entangled parameters. 

Under any reasonable condition chambers operated at gas gains around 
3000, even using a very high drift field of 10 kV/cm, are severly 
damaged within hours running under HERA-B conditions. This sad 
conclusion we verified experimentally in December 1996. To survive 5 
years the discharge rate has to be reduced by 4 orders of magnitude, 
resulting in a gas gain below 1000 and a marginal efficiency of the 
device. 
\\[0.3cm]
In view of this universal nature of the induced discharge phenomenon 
we tried to find out if the tolerance of the chamber to discharges 
can be positively effected by using a clever strip material. In fact 
we learned very soon, that gold strips are extremely delicate whereas 
chromium strips tolerate an enormous number of sparks before showing 
visible marks or even anode breaks  \cite{kell}. 

For gold strips on the other hand the charge stored in the capacity 
of a single anode-cathode system with 10 cm long strips is sufficient 
to damage the anode in a single discharge severely. Unfortunately, 
the resistivity of chromium is that high that the signal risetime for 
strips longer than a few centimeters become unacceptably long for a 
fast detector. After troublesome experiments with electrodes made 
from gold, chromium, aluminum, rhodium  \cite{cbr} and tungsten  
\cite{kell} we had to conclude that it is the mere resistivity and 
resulting current limitation what protects the electrodes from being 
damaged. Spark tolerance and fast signals are therefore incompatible 
requirements. 

\section{The GEM-MSGC}
The recovery of the HERA-B MSGC tracker came with the introduction of 
the Gas Electron Multiplier (GEM) by F. Sauli  \cite{sauli2}. The 
basic idea is to separate the total gas gain in two independent 
factors both sufficiently small to strongly suppress induced 
discharges. The initial major concerns against the technology of the 
GEM-MSGC where as follows:\\
\begin{itemize}
\item Does it really solve the induced discharge problem or is the 
total achievable gain now limited by induced GEM- or combined 
GEM-MSGC discharges? 
\item How is the efficiency, the strip multiplicity and the 
resolution of such a device especially if operated in a magnetic 
field? 
\item What is the long term stability of the GEM in view of aging and 
the negative experience with uncoated glass plates? 
\end{itemize}
Since March 1997 we tried to answer these questions in fruitful 
collaboration with F. Sauli and his team at CERN. The most 
fundamental answer we got rather quickly: \\[0.3cm]
The GEM-MSGC showed now induced discharges under the combined action 
of the gaseous-a source and an x-ray charge load of 20 times HERA-B 
conditions, even when operated at total gas gains above 4000. After 
troublefree operation of a prototype chamber at the HERA-B beamline 
with full interaction rate and a total gas gain of 3000 for more than 
58 hours we came to the conclusion that the problem of induced 
discharges is solved by using a GEM-MSGC combination.
\\[0.2cm]
The second concern could be settled using the electron test beam at 
DESY in July 1997. As shown in figure 11, the efficiency of the 
GEM-MSGC is excellent and completely uneffected by a magnetic field 
of 0.85 T parallel to the strips. Operated with Ar/DME 50:50 the 
strip multiplicity at 95 \% efficiency is 1.65 for 300 $\mu$m wide 
strips, only slightly higher than without the GEM (1.4).

The long term behaviour and aging properties of the GEM-MSGC was one 
of the main concerns which finally could be dispeled by pain staking 
tests at Heidelberg and CERN. The problem was aggravated by the fact 
that not only the GEM introduced new materials in the chamber but 
also the more robust frame replacing our initial glass tube design. 
All GEM's produced on polyimid foil exhibit time dependend and local 
gain variations whose amplitude depends on the details of the GEM 
geometry. Part of these variations are due to surface charge and can 
be avoided by adding a small amount of water to the counting gas. 
Another part has to do with polarization of the polyimid foil and is 
unaffected by the gas humidity. 

Unfortunately adding water to the gas negatively affects the aging 
properties at the MSGC surface. With 3000 ppm of water we observed a 
very rapid degradation of the MSGC gain which turned out to be due to 
deposits on the anode strips. Both, Ar/DME and Ar/CO$_2$ mixtures 
behaved very similarly and thus excluded the use of water admixture 
to reduce surface charge. 

In figure 12 the behaviour of otherwise identical chambers with and 
without water admixture are confronted. 

The entire gain history of a GEM-MSGC running with Ar/DME up to an 
accumulated rate corresponding to 3.5 years of HERA-B operation at 
gain 3000 is shown in figure 13. The initial gain excursion by a 
factor of 1.5 is clearly seen as well as the stabilisation after a 
few days of operation. After the initial period the gain is constant 
even if the chamber is switched off and repowered after several hours.

The GEM-MSGC obviously is a device that can be expected to run 
reliably under HERA-B conditions with high rates for several years. 
Even if the initial gain variations as well as the local fluctuations 
of the GEM amplification factor are no fundamental problem for the 
envisaged HERA-B tracker, they are a drawback making the operation of 
the chamber and the definition of the trigger threshold more 
delicate. 

Recently it has been shown by the HERA-B group at Siegen that both 
effects can be completely avoided by overcoating the GEM with a high 
resistive layer of amorphous carbon using the same plasma technology 
as for the MSGC plate. 
If this technology can be successfully applied to the sizes as needed 
for the HERA-B chambers, it would further enhance the performance of 
the detectors. 

\section{Passivation of strip ends}
Passivation of the strip ends is usually done by coating this 
dangerous area by insulating glue either inside or outside the 
counting gas volume. In our initial design the passivation was 
combined with the gluing of the frame on the MSGC plate. With this 
technology it is unavoidable that small amounts of glue protrude on 
the MSGC surface leading to the situation shown in figure 14 with an 
insulating layer on top of the electrodes inside the counting gas. 
For whatever reasons such topping insulators are introduced, they are 
a source of potential severe trouble.

In figure 15 we show the time dependent calculation of the electric 
field under these conditions. The insulator surface is charged until 
no more field lines ending there. By this very dangerous hot spots 
right after the edge of the insulator are created. At these points 
the chamber is prone to discharges in radiation fields even at very 
moderate gains below 1000. 

Fortunately the problem can be cured in a very elegant way, by just 
leaving the strip ends unpassivated without any coating in the 
counting gas volume. We verified experimentally that free, properly 
designed, strip ends will not cause any trouble even under the 
combined action of the gaseous $\alpha$-source and a very heavy x-ray 
load of several times HERA-B conditions. This positive result could 
be confirmed in the high-intensity pion beam at PSI in Switzerland. 
Avoiding the notorious strip end passivation strongly simplifies the 
chamber construction and reduces the demands on the glue and the 
gluing procedure.

\section{Status and prospects of chamber production}
The MSGC plates for the HERA-B inner tracker are designed by the 
HERA-B group at the University of Z"urich and produced at IMT in 
Greifensee, Switzerland. 
Meanwhile a first batch of 40 plates for the 1998 preseries has been 
produced. The diamond coating was done at Fraunhofer-Institut f"ur 
Schicht- und Oberfl"achentechnik, Braunschweig. These coatings are of 
very good homogenity and surface quality. All GEM's have been 
designed by Fabio Sauli and produced at the CERN workshop. Now in 
January 1998, we have started the production of a preseries of 
chambers which will be installed for the 1998 running of HERA-B. 
During this year the mass production of the full set of about 200 
chambers is foreseen. Installation and commissioning of the chambers 
will take place in the winter shut-down of HERA 1998/99.

\section{Figures}
\setlength{\footskip}{4cm}
\begin{center}
\begin{figure}[ht]
\epsfig{file=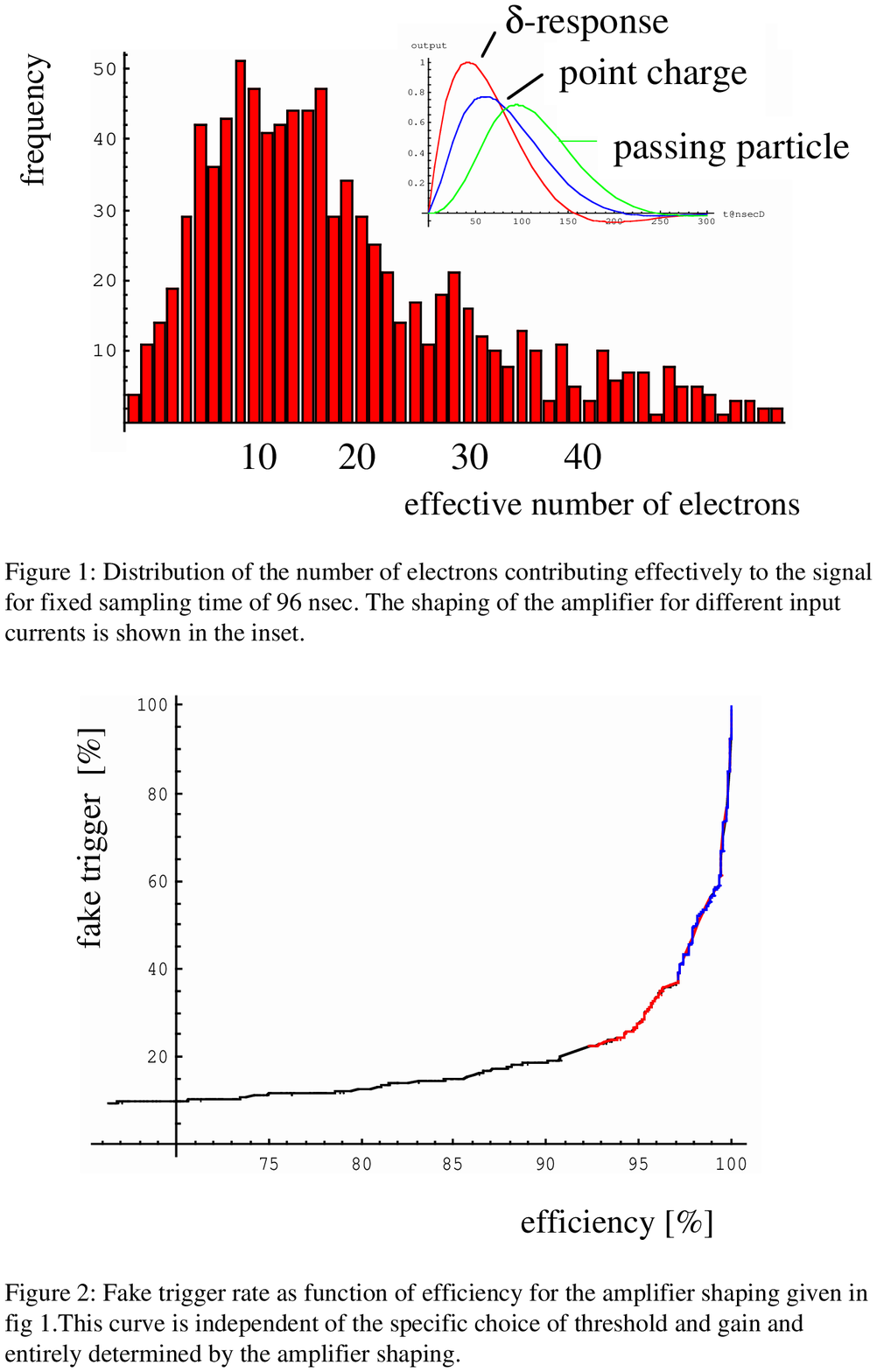,scale=0.7,angle=0}
\end{figure}
\begin{figure}[ht]
\epsfig{file=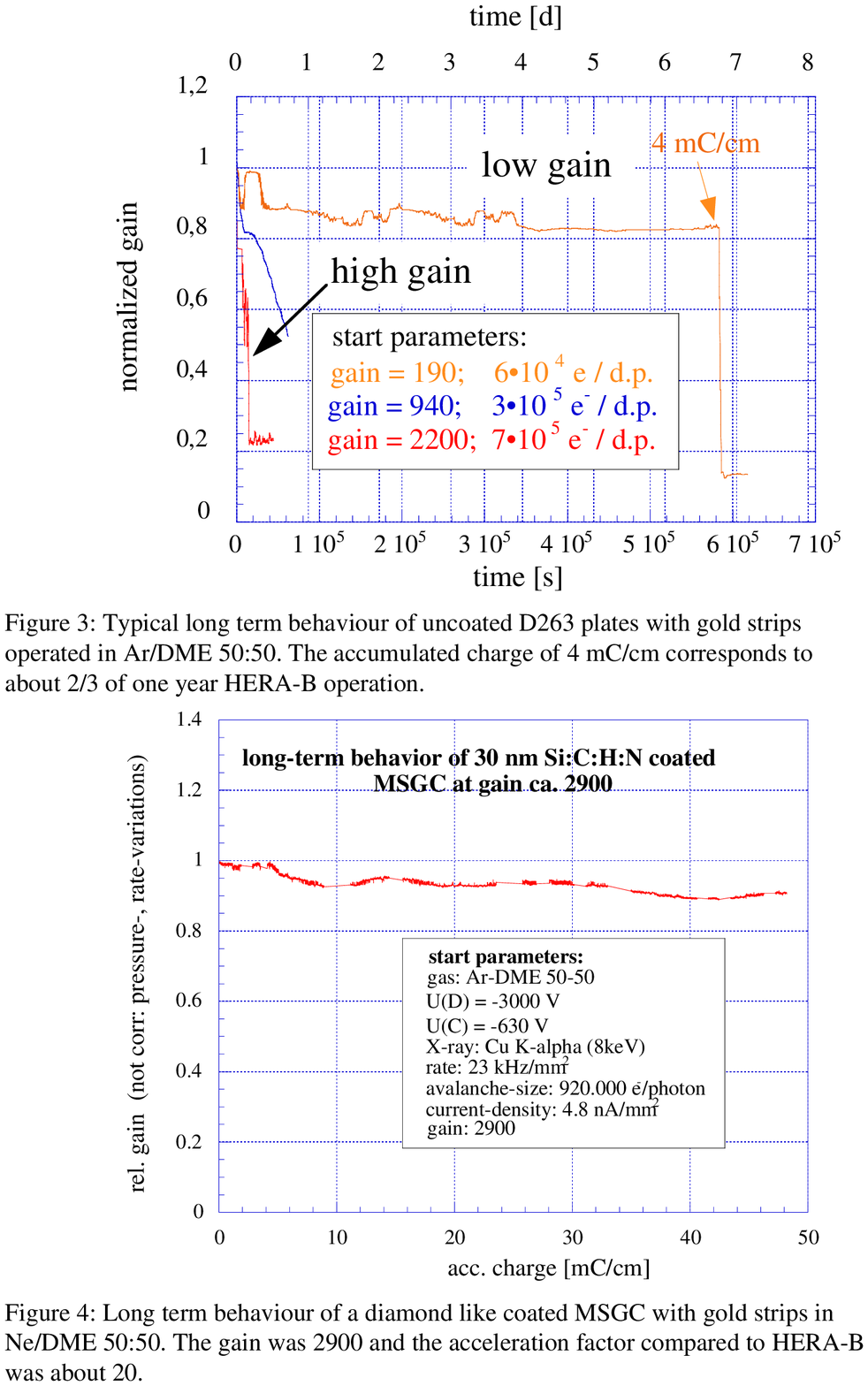,scale=0.7,angle=0}
\end{figure}
\begin{figure}[ht]
\epsfig{file=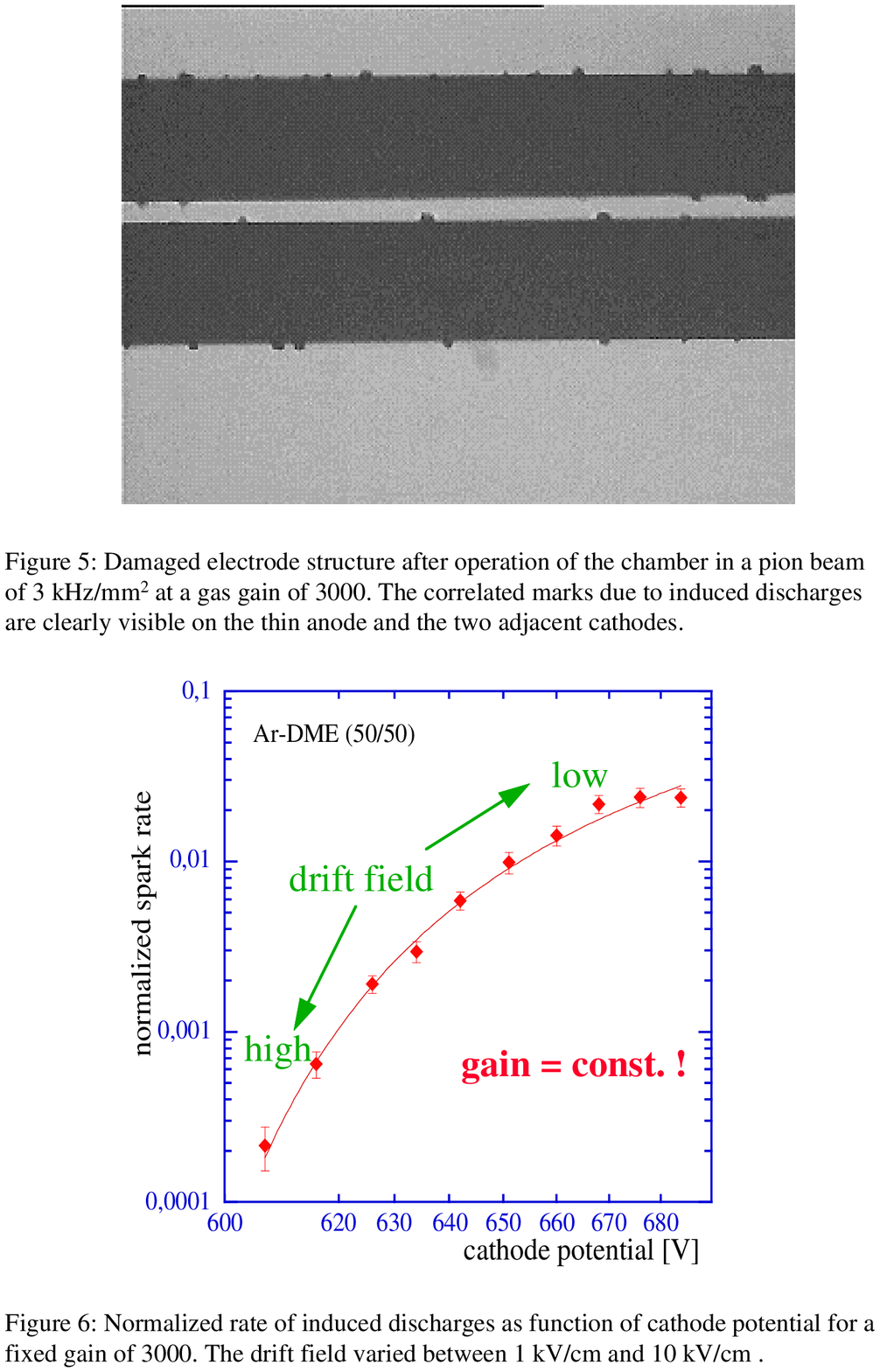,scale=0.7,angle=0}
\end{figure}
\begin{figure}[ht]
\epsfig{file=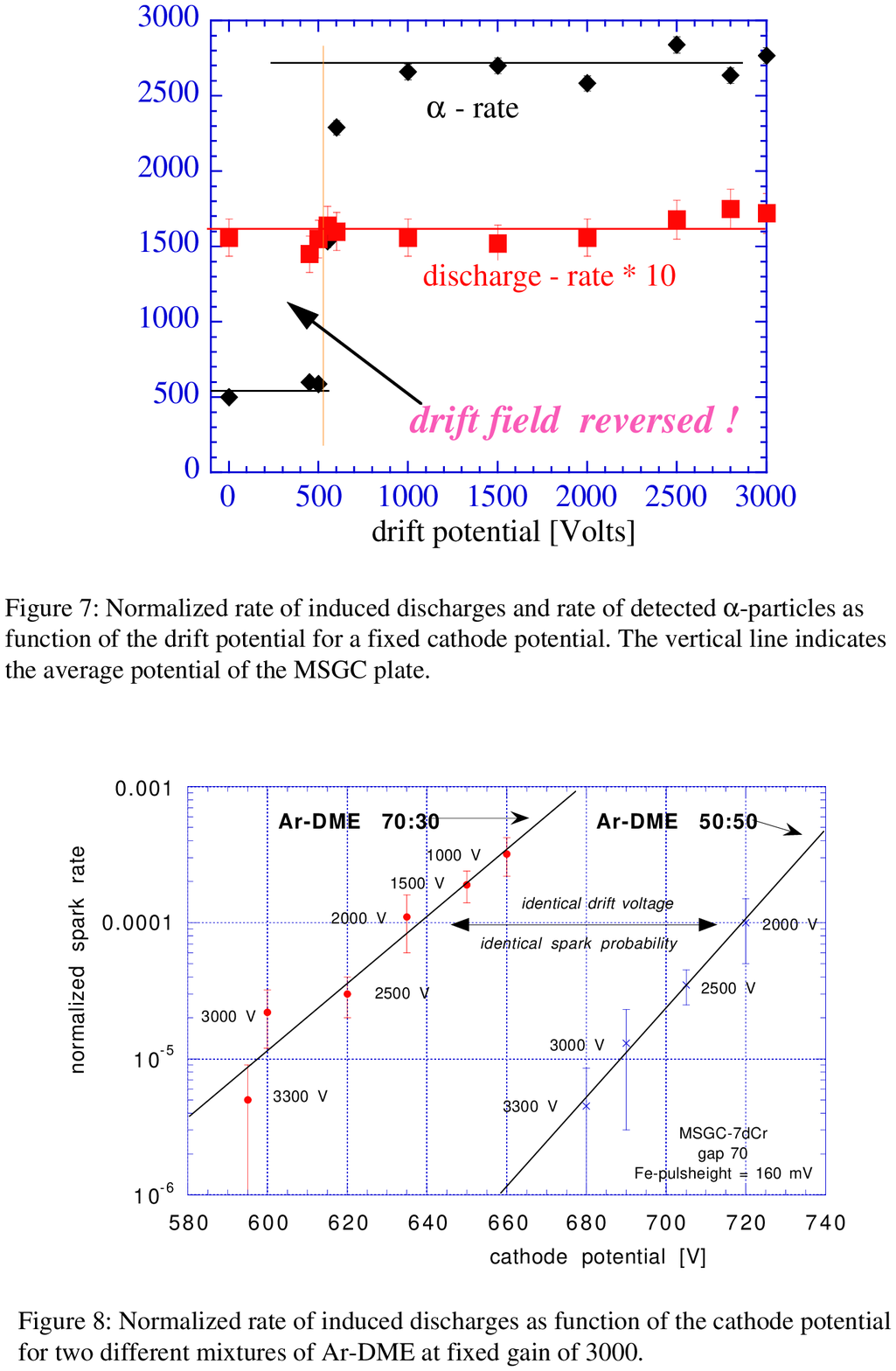,scale=0.7,angle=0}
\end{figure}
\begin{figure}[ht]
\epsfig{file=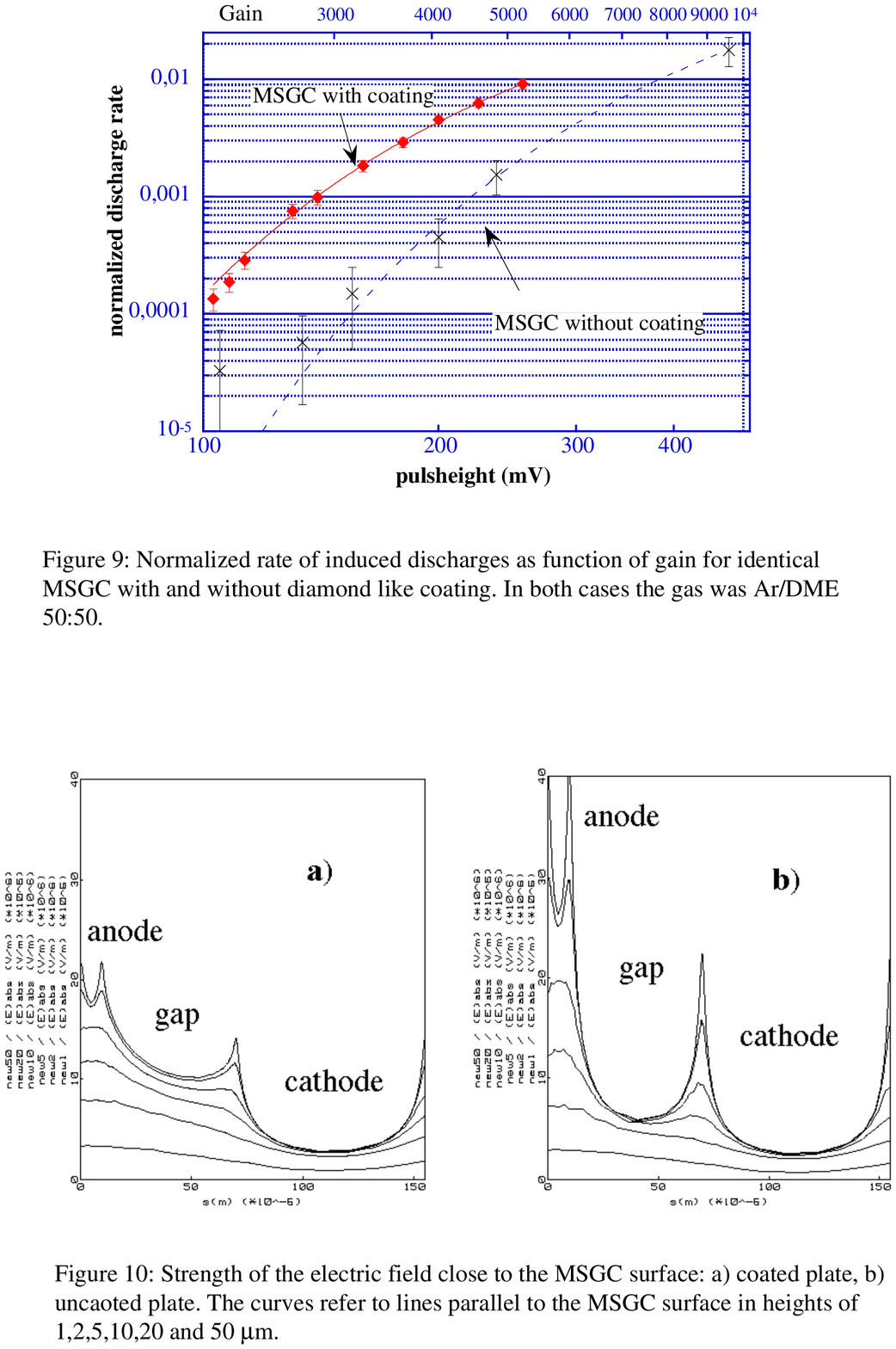,scale=0.7,angle=0}
\end{figure}
\begin{figure}[ht]
\epsfig{file=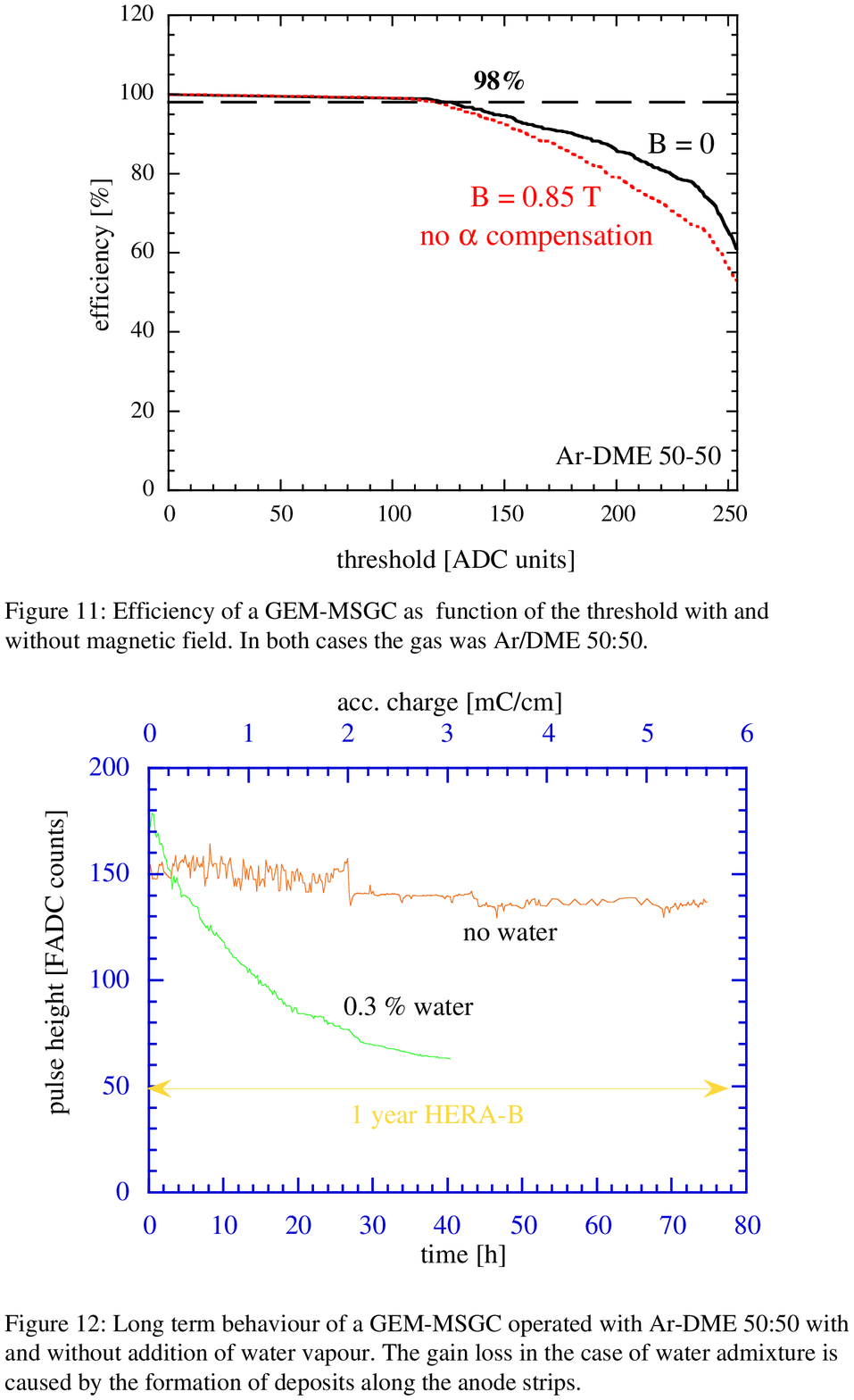,scale=0.7,angle=0}
\end{figure}
\begin{figure}[ht]
\epsfig{file=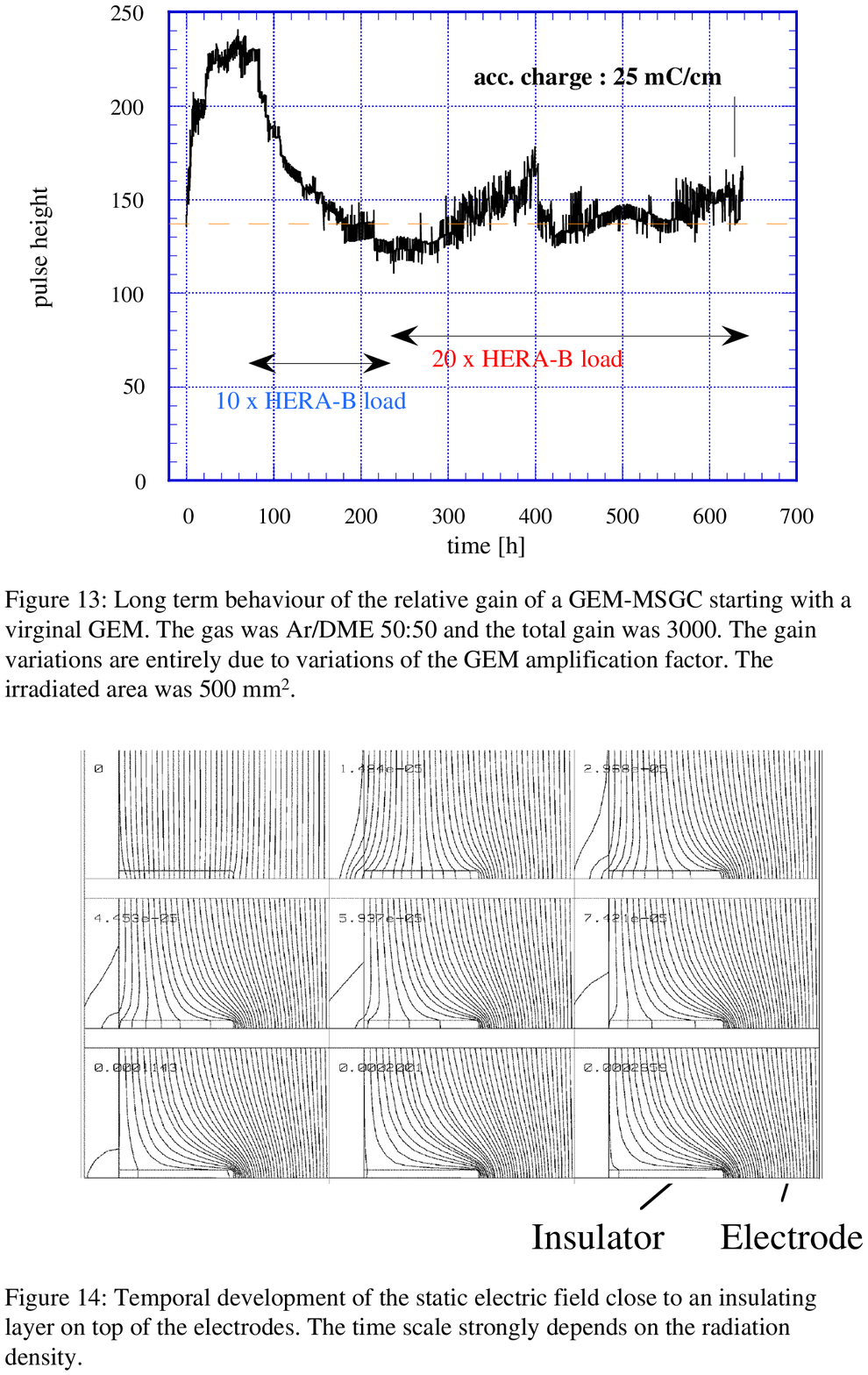,scale=0.7,angle=0}
\end{figure}
\begin{figure}[ht]
\epsfig{file=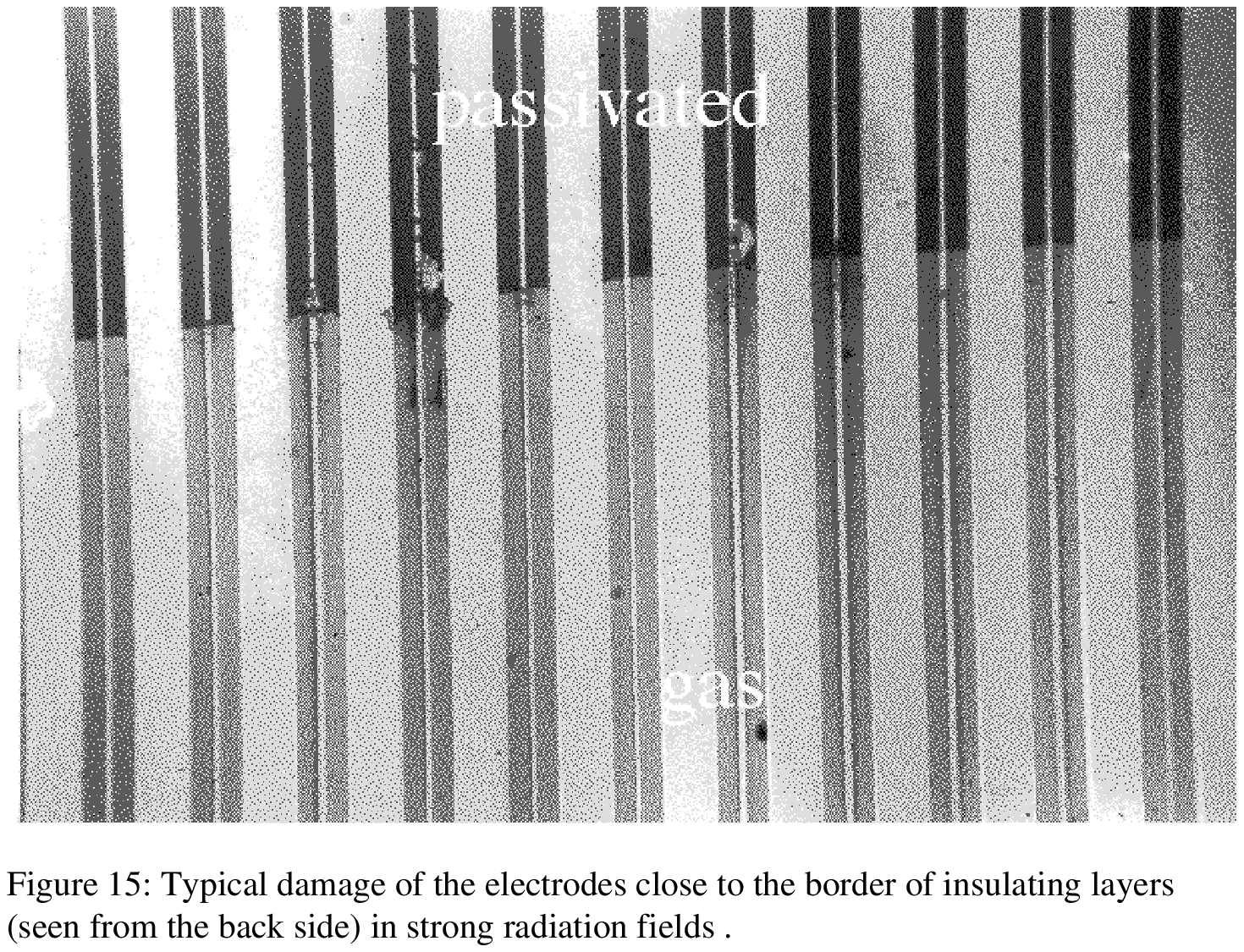,scale=0.7,angle=0}
\end{figure}
\end{center}

\end{document}